\documentclass[
    ,final           
  ]{aipproc}

\layoutstyle{6x9}

\usepackage{epsfig,natbib}
\usepackage{tabularx,array,booktabs,calc,multirow,colortbl,amsmath}
\begin{document}

\title{Hadronic Molecules and the $f_0(980)$/$a_0(980)$}


\author{Tanja Branz}{
  address={Institut f\"ur Theoretische Physik,
Universit\"at T\"ubingen,
\\ Auf der Morgenstelle 14, D-72076 T\"ubingen, Germany}
}

\author{Thomas Gutsche}{
  address={Institut f\"ur Theoretische Physik,
Universit\"at T\"ubingen,
\\ Auf der Morgenstelle 14, D-72076 T\"ubingen, Germany}
}

\author{Valery E. Lyubovitskij}{
  address={Institut f\"ur Theoretische Physik,
Universit\"at T\"ubingen,
\\ Auf der Morgenstelle 14, D-72076 T\"ubingen, Germany}
}

\begin{abstract}
We discuss a possible interpretation of the $f_0(980)$ and $a_0(980)$ mesons 
as hadronic molecules - bound states of $K$ and $\bar K$ 
mesons. Using a phenomenological Lagrangian approach we calculate 
the strong $f_0(980) \to \pi\pi$ and $a_0(980) \to \pi\eta$ as well as the electromagnetic 
$f_0(980) \to \gamma\gamma$ and $a_0(980) \to \gamma\gamma$ decays. The covariant and gauge invariant model, which also allows for finite size effects of the hadronic molecule, delivers results in good agreement with available data and results of other theoretical approaches. 
 \end{abstract}

\maketitle
\section{Introduction}

The scalar $f_0(980)$ and $a_0(980)$ mesons have been analyzed in various structure interpretations amongst the most important are: quarkonium state, tetraquark configuration and hadronic molecule. In particular, the closeness to the $K\bar K$ threshold and their near mass degeneracy, problematic in the $q \bar q$ picture, give evidence for a hadronic bound state interpretation. In addition, recent calculations based on QCD sum rules and lattice QCD also support the $q^2\bar q^2$ configuration, where the quarks can either form a compact tetraquark or a loosely bound state of kaons \cite{Liu:2007hma,Chen:2007mp}.

We present a clear and straightforward model which allows for a consistent evaluation of electromagnetic and strong decay properties of the $a_0$ and $f_0$ considered as pure $K\bar K$ bound states \cite{Branz:2007xp}. Covariance and gauge invariance are the main features of our theoretical framework which additionally considers the spatially extended structure of the hadronic molecules with a minimal amount of assumptions.

\section{Setup of the model}

In this section we focus on the 'supporting pillars' of our framework. The model is based on an interaction Lagrangian describing the coupling between $f_0$ and its $K\bar K$ constituents
\vspace{-0.1cm}
\begin{equation}
{\cal L}_{f_0K\bar K}(x)=g_{f_0K\bar K}f_0(x)\int dy \,\;\Phi(y^2)\bar K\Big(x-\frac y2\Big)K\Big(x+\frac y2\Big)\,,\label{eq:L}
\end{equation}
\vspace{-0.1cm}
with a similar expression for $a_0$.
Here, ${\cal L}_{f_0K\bar K}(x)$ is expressed by the center-of-mass coordinate x and the relative coordinate y. The compositeness condition \cite{Weinberg:1962hj,Scadron:1997nc} provides a self-consistent method to fix the coupling $g_{f_0K\bar K}$ between the $f_0$ bound state and its constituents. 
In order to describe the bound state of constituents the field renormalization constant $Z_{f_0}$ is set to zero
\vspace{-0.2cm}
$$
Z_{f_0}=1-g_{f_0K\bar K}^2\Pi^\prime (m_{f_0}^2)=0\,,
\vspace{-0.1cm}
$$
where $g_{f_0K\bar K}^2\Pi(m_{f_0})$ is the mass operator.

The correlation function $\Phi(y^2)$ in (\ref{eq:L}) allows to account for the finite size of the $f_0$ as a bound state of $K\bar K$. Although the vertex function is related to the shape and size of the meson its explicit form only plays a minor role. The second task of the form factor is the regularization of the kaon loop integral. Here we have chosen a Gaussian form 
\vspace{-0.1cm}
$$
\Phi(y^2)=\int\frac{d^4k}{(2\pi)^4}e^{-iky}\widetilde \Phi(-k^2),\quad\widetilde\Phi(k_E^2)=\exp(-k_E^2/\Lambda^2),
$$ 
\vspace{-0.1cm}
where the index $E$ refers to Euclidean momentum space.

\vspace{-0.2cm}
\section{The electromagnetic decays}

In this section we study the electromagnetic decays of the $f_0(980)$ and $a_0(980)$, which proceed via the charged kaon loop. We derive the form factors in a manifest gauge-invariant way by evaluating the kaon loop integrals and finally deduce the couplings and decay widths. In the following the radiative decay is discussed for the case of the $f_0$, the $a_0$ is treated in full analogy. 

\begin{figure}[tb]
 \includegraphics[trim= 0cm 0cm 0cm 0.0cm, clip,scale=0.57]{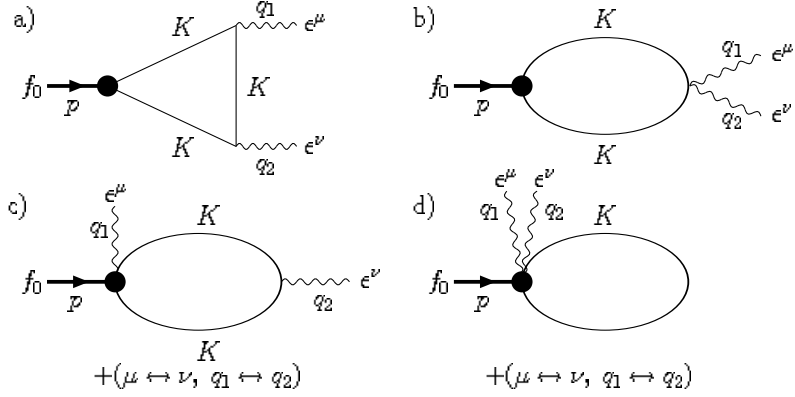}
 \caption{Diagrams contributing to the electromagnetic $f_0\rightarrow\gamma\gamma$ decay.}
\label{fig:1}
\end{figure}

First we restrict to the local case which corresponds to a vertex function with $\lim\limits_{\Lambda\rightarrow \infty}\widetilde\Phi(-k^2)=1$ in the phenomenological Lagrangian (\ref{eq:L}). For this case the electromagnetic fields are included via minimal substitution. The resulting diagrams for the electromagnetic decay are figured in Figs. \ref{fig:1} a) and b).

As mentioned above, the spatially extended structure of the molecule is taken into account by inserting the correlation function $\Phi(y)$. As a consequence gauge invariance of the strong interaction Lagrangian (\ref{eq:L}) gets lost. Hence, we deal with a modified gauge-invariant form 
\begin{eqnarray*}
{\cal L}_{f_0K\bar K}^{GI} &=&g_{f_0K\bar K}f_0(x)\!\!\int\!dy\Phi(y^2)\\
&\times&\big[e^{-ieI(x+\frac y2,x-\frac y2)}K^+\big(\textstyle{x+\frac y2}\big)K^-\big(\textstyle{x-\frac y2}\big)
+K^0\big(\textstyle{x+\frac y2}\big)\bar K^0\big(\textstyle{x-\frac y2}\big)\big]
\end{eqnarray*}
which additionally includes photons via the path integral $I(x,y)=\int\limits_y^x\,dz_\mu A^\mu(z)$ \cite{Terning:1991yt}. Diagrammatically, vertices with additional photon lines attached are generated corresponding to the graphs of Fig. \ref{fig:1} c) and d). It is important to note that these diagrams only give a minor contribution to the transition amplitude but are required in order to fully restore gauge invariance.
\vspace{-0.5cm}
\subsection{Results}
Our results for the $f_0\rightarrow\gamma\gamma$ decay width are ($m_{f_0}$=0.98 GeV)
\vspace{-0.2cm}
\begin{eqnarray*}
\Gamma_{f_0\rightarrow\gamma\gamma}=0.25\text{ keV}\quad(\Lambda=1\text{ GeV})\quad \text{ and }\quad\Gamma_{f_0\rightarrow\gamma\gamma}=0.29\text{ keV} \text{ (local)}\,,
\vspace{-0.2cm}
\end{eqnarray*}
which are in good agreement with experimental data:\\\vspace{-0.2cm}
\begin{table}[tbph] 
\begin{tabularx}{\linewidth}{Xcccc}
\toprule\hline
Reference & ~\cite{Yao:2006px}&~\cite{Mori:2006jj}&~\cite{Marsiske:1990hx}&~\cite{Boyer:1990vu}\\\midrule
$\Gamma(f_0\rightarrow\gamma\gamma)$ [keV]&$0.29^{+0.07}_{-0.09}$&$0.205^{+0.095\,+0.147}_{-0.083\,-0.117}$&$0.31\pm0.14\pm0.09$&$0.29\pm0.07\pm0.12$\\\hline\bottomrule
\end{tabularx}
\end{table}
\newline
For the $a_0$ we used $m_{a_0}$=0.9847 GeV and obtain
\vspace{-0.2cm}
\begin{eqnarray*}
\Gamma_{a_0\rightarrow\gamma\gamma}=0.20\text{ keV}\quad(\Lambda=1\text{ GeV})\quad \text{ and }\quad\Gamma_{a_0\rightarrow\gamma\gamma}=0.23 \text{ keV}\text{ (local)}\,,
\vspace{-0.2cm}
\end{eqnarray*}
\begin{figure}[b]
 \includegraphics[angle=-90,scale=0.35]{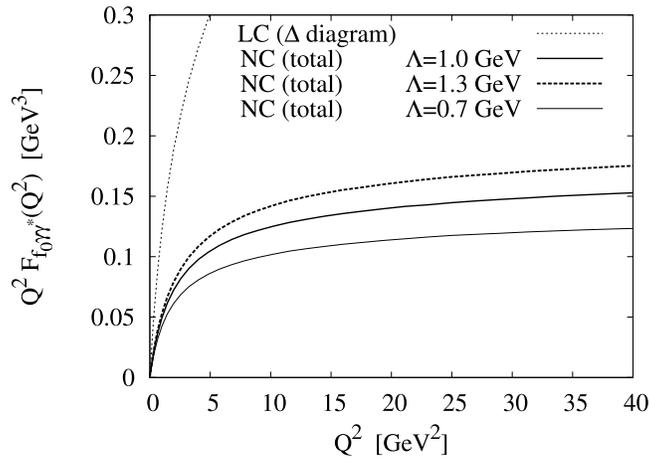}
\label{fig:2}
 \caption{The form factor $Q^2F_{f_0\gamma\gamma^\ast}(Q^2)$ in dependence on $Q^2$ for the local (LC) and nonlocal case (NC). For the latter the form factor is given for different size parameters 0.7, 1.0 and 1.3 GeV.}
\end{figure}
lying within the quoted range of present data $\Gamma_{a_0\gamma\gamma}=0.3\pm0.1$ keV \cite{Amsler:1997up}.
Unfortunately, theoretical predictions of different underlying meson structures (e.g. $q\bar q,\; q^2\bar q^2$) overlap (Tab. \ref{tab:2}). Therefore, at this stage, the radiative decay cannot be used to determine the structure content of the $a_0$ and $f_0$. However, the $K\bar K$ molecular configuration is sufficient to describe the electromagnetic decay.

For the $f_0\rightarrow\gamma\gamma$ decay properties, finite size effects play a minor role when both photons are on-shell. In contrast, the form factor depends strongly on the size parameter $\Lambda$ in case of virtual photons. In Fig. \ref{fig:2} we indicate the form factor $F_{f_0\gamma\gamma^\ast}(Q^2)$ with one real and one virtual photon with Euclidean momentum squared $-Q^2$. To demonstrate the sensitivity of this form factor on finite-size effects we plot the results both for the local case and for the nonlocal vertex function with different size parameters $\Lambda$=0.7, 1.0 and 1.3 GeV. In summary, an experimental determination of $F_{f_0\gamma\gamma^\ast}(Q^2)$ might pose a possibility to identify the underlying structure of the $f_0$/$a_0$.
\begin{table}[t] 
\begin{tabularx}{10cm}{Xccccc}
\toprule\hline
Reference &\cite{Schumacher:2006cy}&\cite{Scadron:2003yg}&\cite{Achasov:1981kh}&\cite{Oller:1997yg}&\cite{Hanhart:2007wa}\\\midrule
Meson structure&$q\bar q$&$q\bar q$&$q^2\bar q^2$&hadronic&hadronic\\\midrule
$\Gamma(f_0\rightarrow\gamma\gamma)$ [keV]&0.33&0.31&0.27&0.2&0.22$\pm$0.07\\\hline\bottomrule
\end{tabularx}
\caption{$f_0\rightarrow \gamma\gamma$ decay width: comparison with $q\bar q,\,q^2\bar q^2$, hadronic approaches.}
\label{tab:2}
\end{table}

\vspace{-0.2cm}
\section{The strong decays}

We also studied the strong $f_0\rightarrow\pi\pi$ and 
$a_0\rightarrow\pi\eta$ decays within our framework. 
Both decays proceed via the diagrams generated by the contact coupling of pions 
and kaons [Fig.3(a)] and $K^\ast$ meson exchange [Fig. 3(b)]. 
Within this model, the $K^\ast$ 
meson is described by antisymmetric tensor fields. As was stressed 
in Ref. \cite{Ecker:1989yg}, the propagators 
$S^V_{K^\ast;\mu\nu,\alpha\beta}(x)$ and 
$S^W_{K^\ast;\mu\nu,\alpha\beta}$ differ by the contact term contained 
in the tensorial propagator
$S^W_{K^\ast;\mu\nu,\alpha\beta}(x)=S^V_{K^\ast;\mu\nu,\alpha\beta}(x)+\frac{i}{m_{K^\ast}^2}[g_{\mu\alpha}g_{\nu\beta}-g_{\mu\beta}g_{\nu\alpha}]\delta^4(x)\,.
$
Using this identity one can show that the contribution of 
the diagram Fig.~3(b) in tensorial representation is given by the sum 
of the graph of Fig.~3(b) in vectorial representation 
plus a graph, which is diagrammatically described by Fig.~3(a), but has 
opposite sign. In addition we obtain a term resulting effectively from 
the difference of two graphs of the type Fig.~3(a), but with different 
numerators in the expression. Numerically it is found that in the last 
term these two contributions almost compensate each other.
\begin{figure}[tbhp]
 \includegraphics[trim= 0cm 0cm 0cm 0cm,clip,scale=0.6]{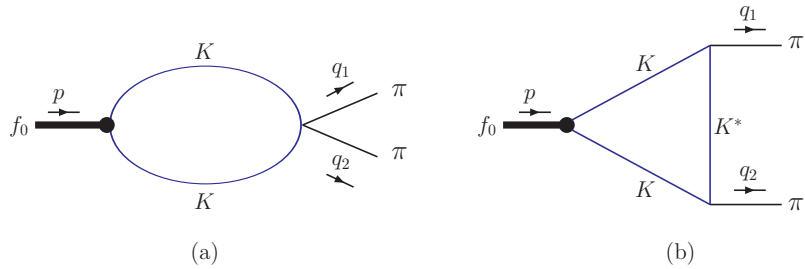}
\caption{Diagrams contributing to the strong $f_0\rightarrow\pi\pi$ decay.}
\label{fig:strong}
\end{figure}
\vspace{-0.5cm}
\subsection{Results}
The experimental results for the dominant decay processes $f_0\rightarrow\pi\pi$ and $a_0\rightarrow\pi\eta$ still have large errors. Our results are compatible with the data as indicated in the following table:

\begin{table}[htpb]
\begin{tabularx}{7cm}{lc}
\toprule\hline
Data &$\Gamma(f_0\rightarrow\pi\pi)$ [MeV]\\\midrule
PDG (2007) (total width)&$40-100$\\\midrule
BELLE ~\cite{Mori:2006jj}&$51.3^{+20.8\,+13.2}_{-17.7\,-3.8}$\\\midrule
Analysis~\cite{Anisovich:2001ay}&$64\pm 8$\\\midrule
Our Result \footnotesize{(isospin limit)}&69 ($\Lambda$=1 GeV) 
\\\hline\bottomrule
\end{tabularx}
\hspace{7pt}
\begin{tabularx}{7cm}{lc}
\toprule\hline
Data &$\Gamma(a_0\rightarrow\pi\eta)$ [MeV]\\\midrule
PDG (2007) (total width)&$50-100$\\\midrule
L3~\cite{Achard:2001uu}&$50\pm13\pm4$\\\midrule
WA102~\cite{Barberis:2000cx}&$61\pm19$\\\midrule
Our Result&59 ($\Lambda$=1 GeV) 
\\\hline\bottomrule
\end{tabularx}
\end{table}

\vspace{-0.6cm}
\section{Summary}
\vspace{-0.1cm}
The scalar $f_0$ and $a_0$ mesons were studied in a hadronic molecule model which is fully covariant and gauge invariant. Additionally, the finite size of the hadronic molecule are taken into consideration which leads to the only free parameter of the model being the cut-off $\Lambda$. Our results are in rather good agreement with experimental data. 

We also showed that for the electromagnetic form factor finite size effects become essential in the case of virtual photons. A more precise measurement of the decay width and in particular an experimental determination of the form factor could help to constrain the $K\bar K$ content of the $f_0$ and $a_0$.
\vspace{-0.3cm}

\begin{theacknowledgments}
\vspace{-0.2cm}
\footnotesize{
This work was supported by the DFG under contracts FA67/31-1 and GRK683. This work is also part of the EU Integrated Infrastructure Initiative Hadronphysics project under contract number RII3-CT-2004-506078 and president grant of Russia ``Scientific Schools'' No. 5103.2006.2.
}
\end{theacknowledgments}
\vspace{-0.2cm}

\bibliographystyle{aipproc} 

\bibliography{lit}

\end{document}